\documentclass[12pt]{article}
\usepackage{graphicx}
\usepackage{cite}

\def \bea{\begin{eqnarray}}
\def \beq{\begin{equation}}

\def \bperp{b_\perp}
\def \bpar{b_\parallel}
\def \eea{\end{eqnarray}}
\def \eeq{\end{equation}}

\def \3half{\frac{3}{2}}

\begin{document}
\begin{flushright}
EFI 10-15 \\
July 2010 \\
\end{flushright}
\centerline{\bf Background check for anomalous like-sign dimuon charge
asymmetry}
\bigskip
\centerline{Michael Gronau}
\medskip
\centerline{\it Physics Department, Technion - Israel Institute of Technology}
\centerline{\it Haifa 32000, Israel}
\medskip
\centerline{Jonathan L. Rosner}
\medskip
\centerline{\it Enrico Fermi Institute and Department of Physics,
 University of Chicago} 
\centerline{\it Chicago, IL 60637, U.S.A.} 
\bigskip
\begin{quote}
The D0 Collaboration has reported an excess of roughly one percent of
$\mu^- \mu^-$ pairs over $\mu^+ \mu^+$ pairs in $\bar p p$ collisions at
a center-of-mass energy $\sqrt{s} = 1.96$ GeV at the Fermilab Tevatron,
when known backgrounds are subtracted.  This excess, if ascribed to
CP violation in meson-antimeson mixing of non-strange or strange neutral
$B$ mesons, is about 40 times that expected in the Standard Model (SM).
We propose a null test, based on a tight restriction on the muon impact
parameter $b$, to confirm that this excess is indeed due to $B$ mesons.
If the asymmetry is due to anomalous CP violation in $B_s$-$\bar B_s$ 
mixing then a tight restriction on $b$ would increase by a factor two
the net asymmetry from neutral $B$ mixing, while the sample of dimuons
from neutral $B$ decays will be reduced significantly relative to
background events.
\end{quote} 

\leftline{\qquad PACS codes:  12.15.Hh, 12.15.Ji, 13.25.Hw, 14.40.Nd} 

\section{Introduction}

The present understanding of CP violation, based on phases in the
Cabibbo-Kobayashi-Maskawa (CKM) matrix, predicts a very small asymmetry in
the yield of same-sign muon pairs due to $b \bar b$ production followed by  
oscillation of a neutral non-strange or strange $B$ meson into its
antiparticle.  Defining
\beq
A^b_{sl} \equiv \frac{N^{++} - N^{--}}{N^{++} + N^{--}}~~,
\eeq
the D0 Collaboration \cite{Abazov:2010hv,Abazov:2010hj} has reported an
asymmetry of
\beq \label{eqn:exptAsl}
A^b_{sl} = -0.00957 \pm 0.00251~({\rm stat}) \pm 0.00146~({\rm syst})~,
\eeq
about a factor of 40 larger than the SM prediction \cite{Lenz:2006hd} of
\beq
A^b_{sl} = (-2.3^{+0.5}_{-0.6}) \times 10^{-4}~.
\eeq
This result is interpreted as evidence at the $3.2 \sigma$ level for anomalous
CP violation in the mixing of neutral $B$ mesons.

The D0 analysis employs a large number of systematic checks to guarantee the
stability of their result.  These 16 checks lead to values of $A_{sl}^b$
consistent with the nominal result (\ref{eqn:exptAsl}).  All these checks used
subsets of data involving about equal fractions of background events
($F_{\rm bkg}$) from non-genuine $b\bar b$ pair production.  To extend such
tests, we suggest a measurement, based in part on a check already performed by
D0, to determine if a substantially smaller asymmetry is obtained in a sample
depleted in $b \bar b$ pairs.  The method employs a reduction of the maximum
allowed impact parameter of muon tracks which should reduce the signal more
than it reduces sources of background.  Basically, we are asking if the D0
experiment can infer zero signal from $b$ decays when no signal is expected.

In Sec.\ II we review kinematics of muons in $B$ decays
and show the effect of changing impact parameter selection criteria.
The relation between impact parameter and $D0's$ selection criteria is
studied in Sec.\ III. Implications of strict impact parameter cuts are 
discussed in Sec.\ IV while Sec.\ V concludes.

\section{Production and decay kinematics}

In $p \bar p$ collisions at the Fermilab Tevatron, $B$ mesons are produced
with a distribution in energy $E$ such that $\beta \gamma = {\cal O}(1)$,
where $\gamma = 1/\sqrt{1-\beta^2} = E/m_B$ (see, e.g., Figs. 2.2 and 2.4 of
Ref. \cite{Anikeev}).  The pseudorapidity $\eta \equiv - \log \tan
(\theta_P/2)$, where $\theta_P$ is the angle with respect to the proton beam,
is typically less than ${\cal O}(3)$ in magnitude.

The average path length traversed by a decaying $B$ meson in its rest frame
is about $c \tau_B = 0.45$ mm, where we take $B^0$ and $B_s$ lifetimes to be
$\tau_B = 1.5$ ps and neglect differences between the two.  Let $t$ denote
the $B$ proper decay time in any event. Then the normalized distribution
in $t$ is $w(t) \equiv (1/\tau) \exp(-t/\tau)$, satisfying $\int_0^\infty w(t)
dt = 1$.  The proper path length $ct$ becomes $\ell = \beta \gamma c t$ in the
laboratory frame.

Define an axis $z^*$ in the $B$ rest frame pointing in the direction of the
boost (with velocity $\beta$) from the laboratory frame to the $B$ rest frame.
Let a muon in $B$ semileptonic decay be emitted with an angle $\theta^*$ with
respect to this axis.  Since the $B$ is spinless, the distribution of the
muon will be isotropic in $\cos \theta^*$.  We seek the transformation from
this angle to the angle $\theta$ in the laboratory frame between the $B$ and
muon directions.  The impact parameter of the muon track with respect to the
primary vertex is then $\ell \sin \theta$, and its distribution is easily
calculated for any value of $\beta$.

\begin{figure}
\begin{center}
\includegraphics[width=0.75\textwidth]{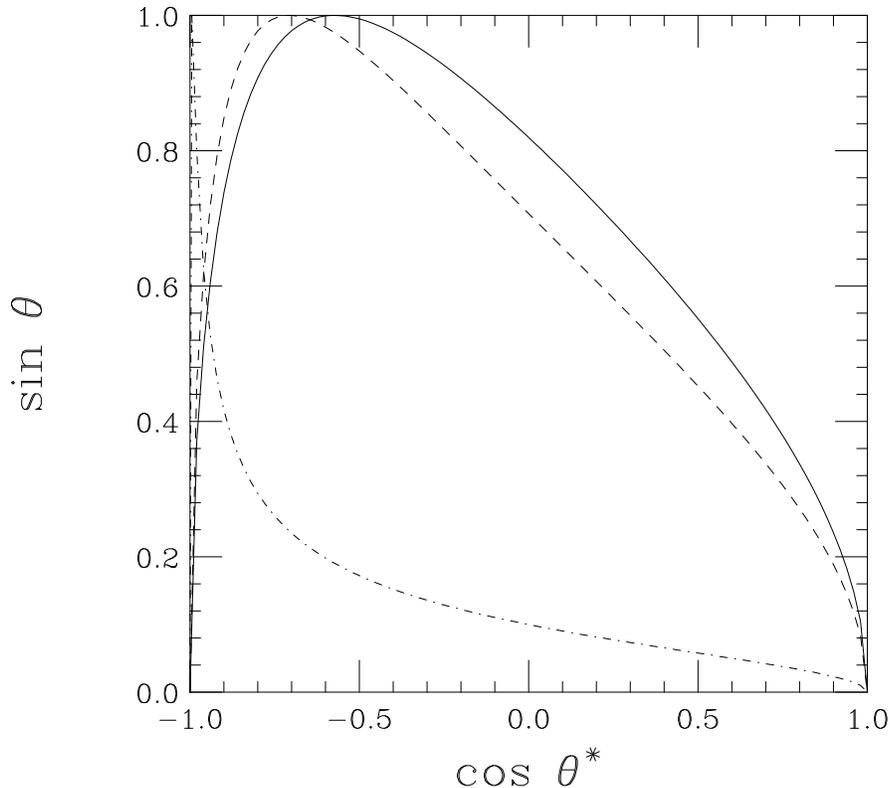}
\end{center}
\caption{Transformation between $\cos \theta^*$ and $\sin \theta$ for
$\beta \gamma = 0.7$ [$\beta = 0.573$] (solid), $\beta \gamma = 1$ [$\beta =
0.707$] (dashed), and $\beta \gamma = 10$ [$\beta = 0.995$] (dot-dashed).
Note that $\theta \to \pi/2$ when $\cos \theta^* = - \beta$.
\label{fig:trans}}
\end{figure}

Define the $x^*$ axis in the $B$ rest frame to be coplanar with the $B$ boost
direction (the $z^*$ axis) and the muon direction.  Then the Lorentz
transformation between momenta in the laboratory frame (unstarred) and the $B$
rest frame (starred) may be written
\beq
p_x = p^*_x~~;~~~
p_z = \gamma(p^*_z + \beta E^*)~~;~~~
E = \gamma(E^* + \beta p^*_z)~~~.
\eeq
Here $E^* = (p^{*2}_x + p^{*2}_z)^{1/2}$, $E = (p_x^2+p_z^2)^{1/2}$;
we have neglected the muon mass.
Then since $p^*_x = E^* \sin \theta^*$, $p^*_z = E^* \cos \theta^*$,
$p_x = E \sin \theta$, $p_z = E \cos \theta$, we have
\beq
\sin \theta = \frac{\sin \theta^*}{\gamma(1 + \beta \cos \theta^*)}~.
\eeq
This reduces, as it should to $\sin \theta = \sin \theta^*$ for $\beta = 0$.
Note that this relation does not depend on the muon energy (in the limit that
$m_\mu$ may be neglected).  We illustrate the transformation for several
illustrative values of $\gamma \beta$ in Fig.\ \ref{fig:trans}.

\begin{figure}
\begin{center}
\includegraphics[width=0.9\textwidth]{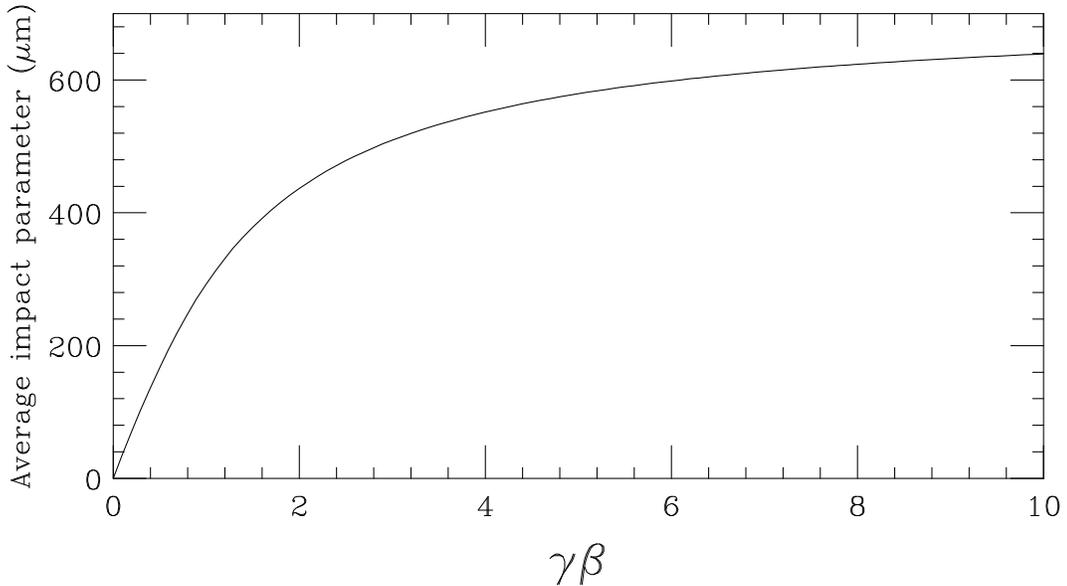}
\end{center}
\caption{Average impact parameter $\langle b \rangle$ as a function of
$\gamma \beta$ for a $B$ decay with assumed $c \tau = 450~\mu{\rm m}$.
\label{fig:impact}}
\end{figure}

Using this transformation and the isotropy of muon emission in the variable
$\cos \theta^*$, one may calculate the average value of $\sin \theta$ and
hence the average impact parameter as functions of $\beta\gamma$:
\beq
\langle b \rangle = \gamma \beta \langle \sin \theta \rangle c \tau~,
\eeq
where
\beq
\langle \sin \theta \rangle = \frac{1}{2}\int_0^\pi \frac{\sin^2 \theta^*~
d \theta^*}{\gamma(1 + \beta \cos \theta^*)} = \frac{\pi}{2}\frac{1}
{1 + \gamma}~.
\eeq
The result is shown in Fig.\ \ref{fig:impact}.

Specifically, for $\gamma \beta = (0.5, 1, 2)$ one has $\langle b \rangle =
(167, 293, 437)~\mu{\rm m}$.
Fitting by eye the exponential tail of an impact parameter distribution
studied by the CDF Collaboration (see Fig.\ 6 of Ref.\ \cite{Aaltonen:2007zza})
we estimate $\langle b \rangle = 350 ~\mu{\rm m}$, which lies in this range.
The impact parameter for a given $\gamma \beta$ will be distributed as 
$W(b) = (1/\langle b \rangle) \exp(-b/\langle b \rangle)$,
where $\int_0^\infty W(b) db = 1$.  If one excludes values $b > b_0$, the
fraction excluded for muons from $B$ decays will be 
$\int_{b_0}^\infty W(b) db = \exp(-b_0/\langle b \rangle)$.  Thus, the remaining 
fraction of dimuons from the two B meson decays will
be $[1 - \exp(-b_0/\langle b \rangle)]^2$. 
We show in Table \ref{tab:frac} the remaining fraction of dimuon events
for various values of $\langle b \rangle$ and $b_0$.
Thus, demanding an impact parameter of less than $100~\mu$m should be enough
to significantly reduce the dimuon signal of two $B$ decays in Refs.\
\cite{Abazov:2010hv,Abazov:2010hj}.  In the next section we discuss the
relation of this criterion to those employed in the D0 analysis.


%
\begin{table}
\caption{Remaining fraction of dimuon events for various values of average
impact parameter $\langle b \rangle$ when events with impact parameter
exceeding a value of $b_0$ are discarded.  Units are $\mu$m.
\label{tab:frac}}
\begin{center}
\begin{tabular}{l c c c c c} \hline \hline
\qquad $b_0$ ($\mu$m) & 100 & 200 & 300 & 400 & 500 \\
$\langle b \rangle$ ($\mu$m) & & & & & \\ \hline
150 & 0.237 & 0.542 & 0.748 & 0.866 & 0.930 \\
300 & 0.080 & 0.237 & 0.400 & 0.542 & 0.658 \\ 
450 & 0.040 & 0.129 & 0.237 & 0.347 & 0.450\\ \hline \hline
\end{tabular}
\end{center}
\end{table}

\section{Relation to criteria employed by D0}

The D0 Collaboration quotes a precision for primary vertex reconstruction
of 20 $\mu$m in the transverse plane and 40 $\mu$m along the beam direction.
Requiring an impact parameter of less than 100 $\mu$m, as proposed above,
thus should not reduce the relative fraction of
signals from tracks originating from the primary
vertex, including kaons, pions, and (anti)protons misidentified as muons
\cite{Abazov:2010hv}.  
However, the employed criteria distinguish between
the transverse impact parameter relative to the closest primary vertex
(which we shall call $\bperp$) and the longitudinal distance from the point
of closest approach to this vertex (which we shall call $\bpar$).  The
D0 selection criteria for these variables are \cite{Abazov:2010hv,%
Abazov:2010hj}
\beq\label{cuts}
\bperp < 3000~\mu{\rm m}~~,~~~
\bpar < 5000~\mu{\rm m}~~.
\eeq
We seek the relation between these quantities and the impact parameter $b$.

Laboratory coordinates $(x,y,z)$ are defined such that the detector axis (very
close to the beam axis) is $z$, the plane of the accelerator is $x$, and the
orthogonal (vertical) direction is $y$ \cite{gbpc}.  The vertex position
$(x_0,y_0,z_0)$ is close, but not necessarily equal, to $x=y=z=0$.  Define
another set of coordinates $(x',y',z')$ for each track such that the
transverse point of closest approach (PCA) has coordinates $x' = \bperp,
y' = z' = 0$, while the vertex is defined to have $x'=y'=0$, $z' = \bpar$.

By definition of the PCA, the linearized approximation to the muon trajectory
near the PCA lies in the $y'$--$z'$ plane, making an angle $\psi$ with the $z'$
axis.  The transverse and longitudinal components of muon momentum in the
laboratory frame, $p_\perp$ and $p_z$, may be used to determine $\psi$:
\beq
p^\mu_\perp = p^\mu \sin \psi~~,~~~
p^\mu_\parallel = p^\mu \cos \psi~~.
\eeq

Let the distance along the muon track from the PCA be denoted by $s$.  The
muon track's coordinates in the primed system are
\beq
y' = s \sin \psi~~,~~~
z' = s \cos \psi~~.
\eeq
The distance $d$ of a point along the muon trajectory from the vertex
satisfies the relation
\beq
d^2 = \bperp^2 + (s \sin \psi)^2 + (s \cos \psi - \bpar)^2~~.
\eeq
We seek the minimum value of this quantity with respect to $s$:
\beq
0 = d(d^2)/ds = 2s - 2 \bpar \cos \psi~~,
\eeq
implying $s = \bpar \cos \psi$ and hence
\beq
d_{\rm min} = [\bperp^2 + (\bpar \sin \psi)^2]^{1/2} = b~~.
\eeq
When $\psi = 0$, $b = \bperp$, while when $\psi = \pi/2$, $b = \sqrt{\bperp^2
+ \bpar^2}$, as expected.  It is then a simple matter to construct the impact
parameter $b$ using the D0 track-selection criteria.

\section{Implication of impact parameter cuts}

Indeed one of the 16 analysis variations mentioned above 
(``Test D''~\cite{Abazov:2010hv}) changes the values of the maximum transverse
and longitudinal impact parameters to values tighter than (\ref{cuts}):
\beq
\bperp < 500~\mu{\rm m}~~,~~~~~~
\bpar < 500~\mu{\rm m}~~.
\eeq
However, since one always has $b \ge \bperp$, one can see from Fig.\
\ref{fig:impact} and Table \ref{tab:frac} that this is not expected to reduce
the signal very much, and it doesn't \cite{Abazov:2010hv}.  While the tighter
cuts reduce the number of like-sign dimuon events by a factor of two, the
fraction of estimated background events is unchanged within errors. (See Table
XIV and Eq.~(37) in Ref.~\cite{Abazov:2010hv}.) 

The much tighter restriction $b \le 100~\mu{\rm m}$ should have a noticeable
effect, reducing significantly the dimuon signal from two muonic $B$ decays
relative to background events.  
The remaining fraction of dimuon signal was calculated in Table I to be 8 and 
4 percent for $\langle b \rangle = 300$ and $450$ microns, respectively.
As a second-order effect, we note that a tight restriction on 
$b$ implies that the charge asymmetry from the remaining neutral $B$ decays 
will be dominated by $B_s$ mesons which involve a much higher oscillation 
frequency than $B^0$. If the anomalous charge asymmetry observed in  
Ref.~\cite{Abazov:2010hv,Abazov:2010hj} originates in $B_s$-$\bar B_s$ mixing 
rather than in $B^0$-$\bar B^0$ mixing then the negative asymmetry from the net
signal events would increase by about a factor of two.

In order to demonstrate this effect, let us take for instance 
$r_b\equiv b_0/\langle b \rangle = 1/3$.  The net asymmetry from
neutral $B$ meson mixing is given by 
\beq
A^b_{sl} = \frac{f_dZ_d}{f_dZ_d + f_sZ_s}\,A^d_{sl} + 
\frac{f_sZ_s}{f_dZ_d + f_sZ_s}\,A^s_{sl}~, 
\eeq
where $A^d_{sl}$ and $A^s_{sl}$ are asymmetries in $B^0$-$\bar B^0$ and 
$B_s$-$\bar B_s$ mixing, respectively. We will take $B^0$ and $B_s$ production 
fractions $f_d=0.323,  f_s=0.118$~\cite{HFAG}, noting that our result will not 
depend on precise values of these parameters.
Neglecting small width differences $(\Delta\Gamma_{d,s}/2\Gamma)^2 \ll1$, 
the parameters $Z_q~(q=d,s)$ are given by
\beq
Z_q = \frac{x^2_q}{1 + x^2_q} - e^{-r_b}\left [ 1 + 
\frac{x_q\sin(x_qr_b) - \cos(x_qr_b)}{1 + x^2_q}\right ]~,
\eeq
where $x_q \equiv \Delta m_q/\Gamma$. Ignoring small errors in neutral $B$
mass differences, $x_d = 0.774, x_s = 26.2$~\cite{HFAG}, 
one finds $Z_d = 0.0029, Z_s = 0.264$ implying $A^b_{sl} = 
0.03A^d_{sl} + 0.97A^s_{sl}$. This result should be compared with 
$A^b_{sl} = 0.51A^d_{sl} + 0.49A^s_{sl}$~\cite{Abazov:2010hv} (also ignoring
errors) obtained with loose restrictions on the muon impact parameter
(i. e., $r_b \gg 1$). Thus, a tight restriction on the impact parameter would 
increase by a factor two the net asymmetry from neutral $B$ mixing if the 
asymmetry originates in the $B_s$ system. 

\bigskip
A tight restriction on impact parameter
eliminates also a large fraction of non-genuine semileptonic $B$ decays. This
includes hadrons in bottom and charm decays, where the hadrons (kaons, pions
and protons) are misidentified as muons, and muons from sequential decays of
kaons and pions. 
In order to provide a null test, or alternatively to demonstrate a nonzero
asymmetry from sources other than $B^0$-$\bar B^0$ and $B_s$-$\bar B_s$ mixing,
the number of remaining events, originating mainly in promptly produced hadrons
faking muons, must be sufficiently large. 
We note that a contribution to the same-sign dimuon asymmetry from kaons,
dominating over those from pions and protons,  must be subtracted from the raw
asymmetry in order to obtain the combined CP asymmetry in $B^0$-$\bar B^0$ and
$B_s$-$\bar B_s$ mixing.  A positive asymmetry of five percent from kaons was
measured in Ref.~\cite{Abazov:2010hv}, involving kaons produced at the primary
vertex and kaons from bottom and charm decays.

\section{Conclusion}

We have proposed a kinematic test to determine if the anomalous like-sign
dimuon charge asymmetry recently reported by the D0 Collaboration
\cite{Abazov:2010hv,Abazov:2010hj} is due to the decays of short-lived
particles such as $B$ mesons.  Although charmed particles are also short-lived,
they are not expected to contribute much to the like-sign signal.  The test
involves a reduction of the maximum impact parameter $b_0$ to a value well
above the vertex resolution but small compared with that characteristic of $B$
decays.  Although we have proposed $b_0 = 100~\mu$m, a dedicated
simulation would probably be preferable in order to optimize this parameter.

\section*{Acknowledgments}

M.G. would like to thank Steven and Priscilla Kersten and the Enrico Fermi
Institute at the University of Chicago for their kind and generous hospitality.
We thank B. Bhattacharya, P. Cooper, G. Giurgiu, J. Lewis, G. Punzi, R.
Tesarek, D. Tonelli, M. Williams and in particular G. Borissov for useful
discussions.  This work was supported in part by the United States Department
of Energy through Grant No.\ DE FG02 90ER40560.

\end{document}